\documentclass[5p]{elsarticle}

\usepackage{amsmath}
\usepackage{amssymb}
\usepackage{amsthm}
\usepackage[usenames, dvipsnames]{color, xcolor}
\usepackage{graphicx}
\usepackage{multirow}
\usepackage{pifont}
\usepackage[caption=false,font=footnotesize]{subfig}
\usepackage{threeparttable}
\usepackage[normalem]{ulem}
\usepackage{url}

\begin{document}

\begin{frontmatter}

\title{A Unified Framework for Wide Area Measurement System Planning}

\author[hku]{James J.Q. Yu\corref{cor}}
\ead{jqyu@eee.hku.hk}
\author[hku]{Albert Y.S. Lam}
\ead{ayslam@eee.hku.hk}
\author[hku]{David J. Hill}
\ead{dhill@eee.hku.hk}
\author[hku]{Victor O.K. Li}
\ead{vli@eee.hku.hk}
\cortext[cor]{Corresponding author}
\address[hku]{Department of Electrical and Electronic Engineering, The University of Hong Kong, Pokfulam, Hong Kong}

\begin{abstract}
Wide area measurement system (WAMS) is one of the essential components in the future power system. To make WAMS construction plans, practical models of the power network observability, reliability, and underlying communication infrastructures need to be considered. To address this challenging problem, in this paper we propose a unified framework for WAMS planning to cover most realistic concerns in the construction process. The framework jointly optimizes the system construction cost, measurement reliability, and volume of synchrophasor data traffic resulting in a multi-objective optimization problem, which provides multiple Pareto optimal solutions to suit different requirements by the utilities. The framework is verified on two IEEE test systems. The simulation results demonstrate the trade-off relationships among the proposed objectives. Moreover, the proposed framework can develop optimal WAMS plans for full observability with minimal cost. This work develops a comprehensive framework for most practical WAMS construction designs.
\end{abstract}

\begin{keyword}
	Wide area measurement system, Dual-Use Line Relays, Phasor Measurement Units, construction, optimization.
\end{keyword}

\end{frontmatter}

\section{Introduction}\label{sec:intro}

Wide Area Measurement System (WAMS), as a reliable monitor of the power network, is considered one of the most important components in the smart grid \cite{ree_synchronized_2010}. In contrast to the current supervisory control and data acquisition (SCADA) system, measurements of the system states are conducted at a much higher rate (5--60 samples per second versus one per 2--6 seconds). In addition, all system phasors are developed simultaneously and continuously, rendering real-time knowledge of power system parameters possible \cite{phadke_state_1986}. As a result, WAMS can significantly improve the performance of power grids by supporting more accurate state estimation, fault detection, stability assessment, remedial control actions, etc. \cite{kamwa_wide-area_2001,fesharaki_simultaneous_2014}.

A typical WAMS comprises synchrophasor measurement devices and Phasor Data Concentrators (PDCs) for aggregating and relaying synchrophasor data. These two major components form a hierarchical structure, connecting through a communication infrastructure (CI). While Phasor Measurement Units (PMUs) are widely employed in WAMS, the recently available Dual-Use Line Relays (DULRs) introduce variability to the modern WAMS construction \cite{pal_pmu_2017}. DULRs are the protection digital relays for transmission lines and transformers which can report synchrophasor data while providing system protection \cite{grigsby_power_2012}. Due to its characteristics of being installed along transmission lines and at transformers, DULR is also called ``branch PMU'' in some previous research \cite{emami_robust_2010,gomez_reliability_2015}. Although DULR can only monitor the voltage phasor of its adjacent bus and the current phasor of the branch, it is still promising due to its low construction cost \cite{pal_pmu_2017}.

Despite much work on WAMS planning, there seems no unified WAMS planning framework that jointly optimizes multiple important objectives simultaneously for placing both measurement devices and PDCs in the system.
\textcolor{blue}{Moreover, despite the decreasing device costs for constructing WAMS, utilities are still a long way from achieving full-installation of PMUs and PDCs across the grid.
Meanwhile, better WAMS construction strategies are still welcomed due to their better system reliability and cost-efficient properties.}

In addition, much previous work suffers from unrealistic assumptions, which have been thoroughly discussed in \cite{pal_pmu_2017}. For instance, while some work considers installing PMUs at buses, they should actually be placed at substations which is a collection of multiple buses. When a PMU or DULR is being installed, the respective substation needs to be interrupted leading to a substantial cost in WAMS construction \cite{_factors_2014}. Consequently a comprehensive model is required to account for all kinds of WAMS installation costs. 

Moreover, as PMUs are generally assumed to be installed on buses in the literature, two buses connected with transformers are both considered observable if either one is equipped with a PMU. However, this hypothesis relies on a model of transformer tap positions as fixed network parameters. The estimated bus voltages, power flows and injections with a transformer with incorrectly modeled or inaccurately measured tap ratio can deviate significantly from their actual values, resulting in inaccurate system state estimation \cite{pal_pmu_2017,korres_transformer_2004}. Last but not least, a majority of previous work assumes that PMUs are equipped with unlimited measurement channels to observe the current phasors of all connecting branches. Other work focuses on minimizing the number of channels required in WAMS \cite{rather_realistic_2015}. However, none of the existing work determines which branches are observed by each PMU. When given optimal branch allocations for PMUs, we can further improve the measurement performance of a WAMS.

In this paper we propose a unified framework, aiming to fulfill different construction requirements for WAMS. The main contributions of this work are listed as follows:
\begin{itemize}
\item We propose a unified framework for WAMS planning, in which PMU, DULR, and PDC placements are jointly optimized simultaneously.
\item We consider a realistic cost model for WAMS construction including the power system substation interruption cost during installation.
\item We consider a practical substation model with unknown transformer tap ratio, which can better facilitate the utilization of measured system synchrophasors.
\item We consider the channel limits of PMUs, and jointly determine which branches should be observed, aiming to provide full observability with the least devices. 
\item Pareto solutions can be determined for decision making considering various requirements of WAMS.
\end{itemize}

The rest of this paper is organized as follows. Section II gives a brief literature review on WAMS construction research. Section III introduces the system model which allows us to design a unified WAMS planning framework. Section IV formulates a multi-objective WAMS planning problem for developing WAMS construction plans. Section V demonstrates the implementations of the proposed framework on IEEE test systems, and compares them with the state-of-the-art solutions. Finally, we conclude this paper in Section VI with discussions on the proposed framework.

\section{Related Work}\label{sec:review}

With the increasing demand of synchrophasor measurement in modern power systems, utilities need a methodology to construct WAMS optimally. Canonically most optimal WAMS construction work focused on finding the minimal number of PMUs to ensure full system observability, subject to pre-defined constraints. This so-called optimal PMU placement problem and its variants have been investigated intensively in the past two decades. A wide range of solution methods have been developed to achieve the optimal solution of this NP-hard problem, including but not limited to integer programming \cite{gou_generalized_2008}, meta-heuristics \cite{milosevic_nondominated_2003}, exhaustive search \cite{chakrabarti_optimal_2008}, weighted least square algorithm \cite{manousakis_weighted_2013}, etc. Interested readers can refer to \cite{manousakis_taxonomy_2012} and \cite{aminifar_synchrophasor_2014} for more details of PMU placement methodologies.

Besides construction cost, system measurement reliability is also critical for WAMS construction. System states of line outages and loss of measurements need to be considered in order to design a robust and reliable WAMS. An intuitive solution to address these system failures is to install duplicate measurement devices to observe the same bus and this is called measurement redundancy. Due to its simplicity, this technique has been widely adopted (see \cite{gou_generalized_2008,aminifar_contingency-constrained_2010} for examples). However this scheme may potentially lead to over-installation of PMUs in the system \cite{pal_pmu_2017}. As an alternative, a reliability-based PMU placement model is proposed where the possibility of maintaining full observability is investigated \cite{gomez_reliability_2015}. This model considers a trade-off between the total number of PMUs installed and the WAMS reliability, resulting in more versatile placement solutions \cite{aminifar_synchrophasor_2014}.

Concurrently optimizing PMU and PDC placement is another research direction related to WAMS planning. \cite{fesharaki_simultaneous_2014} manipulates the placement of measurement devices and PDCs to construct multiple data paths for the generated synchrophasors to overcome CI failures in WAMS. \cite{rather_realistic_2015} and \cite{mohammadi_new_2016} try to minimize the system scale of CI to reduce the WAMS construction cost.

\textcolor{blue}{There is also recent work analyzing WAMS construction from the perspective of graph theory and network equivalency \cite{chakrabortty_introduction_2013}.
While satisfying the conventional bus observability constraint, \cite{anderson_pmu_2014} also enables estimation of system dynamic models by network reduction approaches.
The results can be further employed to update the offline system model.}

\textcolor{blue}{Besides, % above previous work focusing on different formulation of the optimization objective, 
there is also research investigating the integration of WAMS construction with other power system applications and services \cite{chakrabortty_introduction_2013,zima_design_2005}.
For instance, state estimation is among the most important power system applications which can greatly benefit from WAMS.
In \cite{ghiocel_phasor-measurement-based_2014}, the impact of PMU placement plans on the reliability of state estimation subject to data integrity issues is investigated.
The proposed mechanism can also provide estimated system parameters given measurement redundancy.
Other analyses take power stability analysis \cite{chakrabortty_optimal_2014,zhang_admm_2016,zweigle_wide-area_2013} and oscillation monitoring \cite{li_wide-area_2016} into consideration.
Another widely investigated integration considers power system cyber-security.
For passive cyber-attack prevention, constructing WAMS and its protection facilities strategically is widely adopted \cite{liang_review_2017}.
Much research has been conducted in this direction, see  \cite{pasqualetti_attack_2013,almas_vulnerability_2017} for instance.
References \cite{alves_review_2007} and \cite{deng_false_2017} provide thorough surveys on this topic.
Another interesting direction considers communication infrastructure in designing WAMS construction and operation plans.
For instance, through proper software layer design, data communication quality-of-service can be guaranteed \cite{gjermundrod_gridstat:_2009}.
Cloud computing may also greatly contribute to improving the communication and computing network efficiencies \cite{xin_study_2013}.
The above are examples of existing research on applications and extensions of the WAMS construction problem.
They all indicate the significance of optimal WAMS construction strategies.
However, due to the limitations of most previous WAMS construction work introduced in Section \ref{sec:intro}, a generalized and realistic formulation of the WAMS construction problem is required.
%This paper is dedicated to formulate and solve this problem.
}

\section{System Model}\label{sec:model}

\begin{figure}
\includegraphics[width=\linewidth]{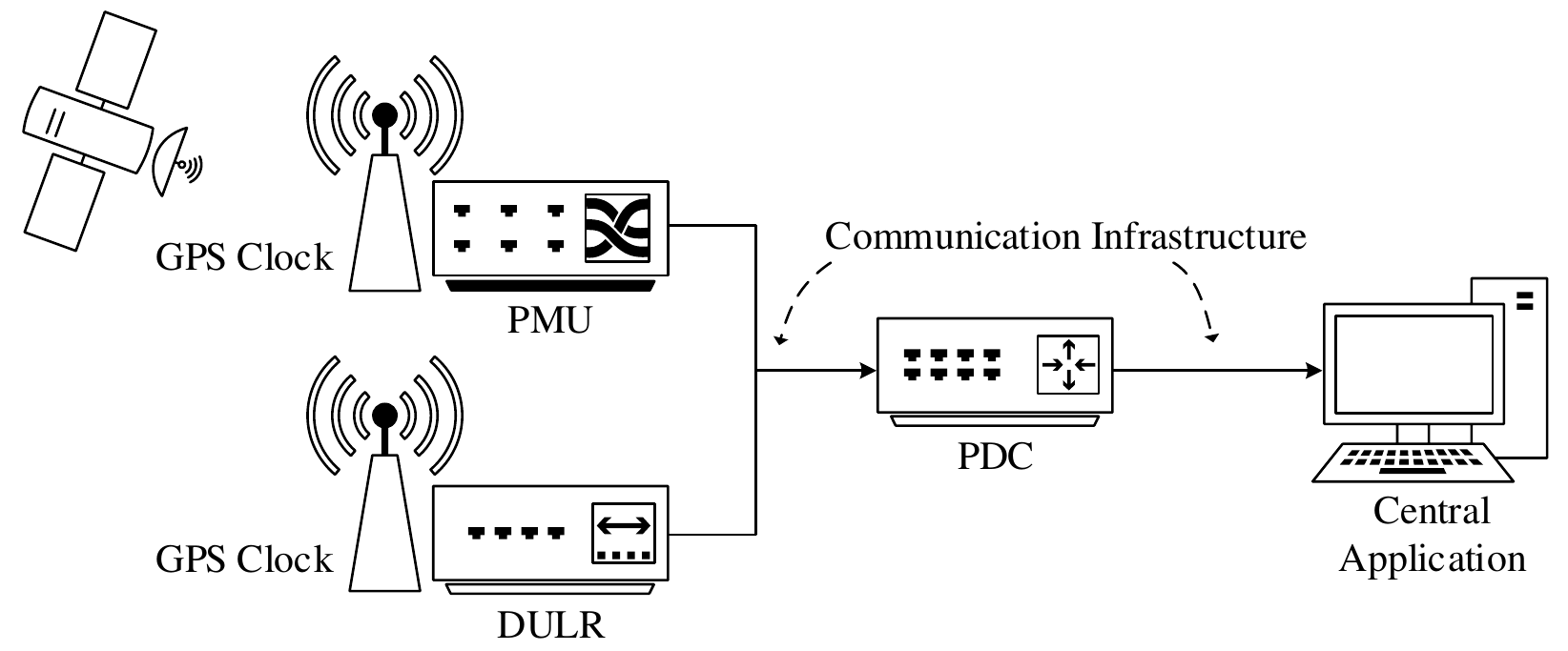}
\caption{Components of a wide area measurement system.}
\label{fig:wams}
\end{figure}

Fig. \ref{fig:wams} depicts a schematic WAMS architecture, where PMU, DULR, PDC, and the central controller form a hierarchical structure over CI, which serves as the media for data transmission. PMU and DULR interface WAMS with the power system and they comprise current transformers (CTs), voltage transformers (VTs), instrumentation cables, and synchronous GPS clocks. Synchrophasors measured by these devices are transmitted to one or multiple layers of PDCs located at selected locations in the system, where the data are aggregated, compressed, and sorted into a time-stamped measurement stream \cite{_ieee_2013}. In general, the data stream is then fed into application software at the central controller for system state monitoring and control decision generation with various control objectives. For simplicity, in this work we assume one layer of PDCs is utilized in WAMS, while multiple layers can be easily adopted into the proposed framework.

\subsection{Power System and Network Observability}\label{sub:observability}

We model the power network with an undirected graph $G(\mathcal{V},\mathcal{E})$, where $\mathcal{V}$ and $\mathcal{E}$ are the sets of buses and branches, respectively. There are $K$ substations, each of which, denoted by $\mathcal{S}_k$, comprises a subset of buses and thus we may simply write $\mathcal{S}_k \subset\mathcal{V}$. For bus $i\in\mathcal{V}$, we denote the set of its neighboring buses by $\mathcal{N}_i=\{j\in\mathcal{V}|(i,j)\in\mathcal{E}\}$.

A power system is deemed observable if the bus voltage phasor (both magnitude and angle) at every bus in the network is known. We consider that maintaining full system observability is one of the critical objectives of WAMS, where the voltage phasor of every bus can either be directly measured or indirectly inferred by calculation. Considering the PMU measurement channel limit, branches associated to $\mathcal{N}_i^\prime\subseteq\mathcal{N}_i$, i.e., $\mathcal{E}_i=\{(i,j)|j\in\mathcal{N}_i^\prime\}$, are observed by a PMU installed on bus $i$. Consequently, the current phasors of these branches and the voltage phasor of $i$ are measured while the voltage phasors of $\mathcal{N}_i^\prime$ can be calculated. This makes buses $i$ and $\mathcal{N}_i^\prime$ observable. Similarly, a DULR placed at the $i$ end of branch $(i,j)$\footnote{For simplicity, we write ``DULR installed on bus $i$'' to represent the DULR placed at the $i$ end of branch $(i,j)$  when no confusion will be caused thereafter.} measures the voltage phasor of $i$ and current phasor of $(i,j)$. This makes the voltage phasor of $j$ calculable. Hence both buses $i$ and $j$ are considered observable.

While buses with measurement devices can have their voltage phasors directly measured, other buses need the current phasors of their connecting branches to infer their voltages. However, these current phasors may not  always be available due to possible line outage contingencies. To specify any  of these contingencies, we define a system state $s$ as a combination of different line outages, where $\mathcal{E}^s\subset\mathcal{E}$ represents the set of failed power transmission lines of system state $s$. Moreover, we describe the bus connectivity  of $s$ with $\mathbf{A}_s=(a_{ij,s})\in\mathbb{B}^{|\mathcal{V}|\times|\mathcal{V}|}$, where $\mathbb{B}=\{0,1\}$ is the Boolean domain. We have $a_{ij,s} = 1$ if $(i,j)\in\mathcal{E}\setminus\mathcal{E}^s$ or $i=j$, otherwise $a_{ij,s} = 0$. For simplicity, $\mathbf{A}=(a_{ij})$ is employed to stand for the bus connectivity of system state $s_0$, where there is no line outage in the system. Furthermore, the bus observability of $s$ is denoted by $\mathbf{O}_s=(o_{i,s})\in\mathbb{B}^{|\mathcal{V}|\times1}$, where $o_{i,s}=1$ means bus $i$ is observable in $s$ and zero otherwise.

\subsection{Phasor Measurement Reliability}\label{sub:reliability}

Designing a robust WAMS for complete system observability with the presence of system failures is critical in WAMS planning. To achieve this, one typical method is to provide measurement redundancy for each bus in the system. That is, we employ multiple measurement devices to provide independent observability to each bus. However, usually more than half of the substations in the system need to be equipped with PMUs  in order to provide full observability under $N-1$ contingencies, but such over-installation is not practical \cite{pal_pmu_2017}. Thus we do not consider full observability guarantee under contingencies in this way.

Instead we address system observability in the probability sense. In practice, two types of failures are commonly considered in WAMS design, namely, line outages and measurement device failures. In this work, both kinds of failure are considered. We address the line outage failures by minimizing the number of unobservable buses for different system states to improve the system reliability. We also provide multiple measurement devices to observe critical buses to minimize the impact of measurement device failures on the whole system, as will be described in Section \ref{sss:observability}.

Each system state $s$ with line outages has a probability of occurrence calculated by
\begin{equation}
p_s=\prod_{l\in\mathcal{E}\setminus\mathcal{E}^s}p_l\prod_{l\in\mathcal{E}^s}(1-p_l),
\end{equation}
where $p_l$ is the reliability of power line $l$. With the measurement devices fixed in the system, the observability of each bus can be developed through considering failed lines in $s$. We define the unreliability index of each bus subject to current measurement device allocation as
\begin{equation}\label{eqn:ui}
U_i=\sum_{s\in S}p_s(1-o_{i,s}),
\end{equation}
where $S$ is the set of all system states. We further define the overall system unreliability index by aggregating the indices of all buses in the system, i.e., 
\begin{equation}\label{eqn:uwams}
U^\text{WAMS}=\sum_{i\in\mathcal{V}}U_i.
\end{equation}

\subsection{Communication Network}\label{sub:communication}

PMU and DULR devices measure the bus voltage and current phasors of their attached branches. Each measurement device generates a data stream to be transmitted to a specific PDC for data aggregation, which then passes the compressed synchrophasor data to the central controller for certain power applications. To reduce the possible data transmission delay due to communication link congestion and data queuing, we should carefully place PMUs and DULRs on the power network in order to reduce the total data traffic transmitted from measurement devices to both PDCs and the central controller.

According to the IEEE Standard for Synchrophasor Data Transfer for Power Systems \cite{_ieee_2011}, the maximum data rate of traffic generated by measurement devices at bus $i$, can be estimated as $(|\mathcal{N}_i|+1)LF$ bit-per-second, where $L$ and $F$ are the data size of each synchrophasor data message and synchrophasor sampling rate\footnote{Here the data frame overhead is omitted as it is normally very small when compared with the synchrophasor data message body.}, respectively. Here current phasors of $|\mathcal{N}_i|$ adjacent branch and one bus voltage phasor give $|\mathcal{N}_i|+1$ sets of measurement data.

After these synchrophasors are generated, they are sent to one dedicated PDC for data aggregation, compression, and relaying. Then the processed data is effectively compressed with any appropriate data compression schemes, e.g., Exception Compression and Swing Door Trending Compression \cite{zhang_application_2015}, and later relayed to the central controller for data utilization. These two kinds of data transmission are achieved through CI, which is in turn supported by an existing telecommunication network.
\textcolor{blue}{We do not impose any limitations on the topology of CI. We assume that all substations are connected through CI, which can be configured with any reasonable topology \cite{torres_communication_2013}.}
We employ the distance matrix $\mathbf{Q}=(q_{ij})\in\mathbb{Z}^{|\mathcal{V}|\times|\mathcal{V}|}$ to describe CI, where $q_{ij}$ is the number of  hops on the shortest path from $i$ to $j$ over CI. Therefore, the maximum data traffic originated from bus $i$ is $(|\mathcal{N}_i|+1)LFH_i$ bit-hop-per-second, where $H_i$ is the number of hops of the shortest path from bus $i$ to any PDCs over graph $G$. Similarly, the data traffic from PDC to the central controller is defined as $\eta N_i^\text{PDC}H_i^\text{PDC}$, where $\eta$ is the data compression ratio, $N_i^\text{PDC}$ is the aggregated data rate received by PDC installed on $i$, and $H_i^\text{PDC}$ is the number of hops of the shortest path from bus $i$ to the central controller.

\section{WAMS Planning Problem}\label{sec:problem}

Based on the models discussed in Section \ref{sec:model}, we formulate a multi-objective realistic optimal WAMS planning problem (WPP) in this section. The objective of WPP is to optimally place and set up PMUs, DULRs, and PDCs in the power system such that
\begin{itemize}
\item the total WAMS construction cost is minimized;
\item the overall system unreliability index is minimized;
\item the total WAMS synchrophasor data traffic from measurement devices to PDCs is minimized;
\item all buses are observable with unknown transformer tap ratio; and
\item all system constraints are satisfied.
\end{itemize}

\subsection{Control and Ancillary Variables}\label{sub:variable}

To formulate the problem, we define several control variables, namely, $\mathbf{M}$, $\mathbf{D}$, $\mathbf{P}$, and $\mathbf{L}$. $\mathbf{M}=(m_{ij})\in\mathbb{B}^{|\mathcal{V}|\times|\mathcal{V}|}$ is the PMU installation and observation matrix, and it determines whether PMUs are installed on specific buses, and whether the current phasors of the respective connected branches are measured. Specifically, if a PMU is installed on bus $i$ implying that the voltage phasor of $i$ is measured, then the diagonal value $m_{ii}$ will be set to one. Moreover, if the current phasor of a branch $(i,j)$ is measured, the corresponding non-diagonal $m_{ij}$ value will also be set to one.

Similarly, $\mathbf{D}=(d_{ij})\in\mathbb{B}^{|\mathcal{V}|\times|\mathcal{V}|}$ is the DULR installation matrix, and if we have $d_{ij}=1$,  a DULR will be installed on branch $(i,j)$ on the $i$ end. However, unlike $\mathbb{M}$, we always set all diagonal values $d_{ii}$'s to zero and this will make the subsequent formulation simpler.

We also let $\mathbf{P}=(p_i)\in\mathbb{B}^{|\mathcal{V}|\times1}$ to denote the PDC installation status, where $p_i=1$ indicates that a PDC is installed on bus $i$. Finally, $\mathbf{L}=(l_{ij})\in\mathbb{B}^{|\mathcal{V}|\times|\mathcal{V}|}$ is introduced to allocate each PMU and its associated DULR devices to a PDC. If we set $l_{ij} = 1$, PMU and all DULRs installed on bus $i$ will be assigned to transmit their measured synchrophasors to the PDC installed on bus $j$.

Besides these control variables, we also define three ancillary variables to facilitate the formulation, including $\mathbf{M}^\prime$, $\mathbf{D}^\prime$, and $\mathbf{U}$. We use $\mathbf{M}^\prime=(m_i^\prime)\in\mathbb{B}^{|\mathcal{V}|\times1}$ as a PMU installation indicator, where $m_i^\prime=1$ if a PMU is installed at bus $i$. In addition, $\mathbf{D}^\prime=(d_i^\prime)\in\mathbb{B}^{|\mathcal{V}|\times1}$ is a measurement device installation indicator, where $d_i^\prime=1$ if either a PMU or a DULR is installed at bus $i$. Meanwhile, when a measurement device is planned to be set up at bus $i$, the associated substation $\mathcal{S}_k\ni i$ needs to be interrupted for installation. Thus we use $\mathbf{U}=(u_i)\in\mathbb{B}^{k\times1}$ to indicate whether each substation will be interrupted, incurring disruption cost. Although $\mathbf{M}^\prime$, $\mathbf{D}^\prime$, and $\mathbf{U}$ can be implied from $\mathbf{M}$ and $\mathbf{D}$, we still adopt these ancillary variables to simplify the problem formulation.

\subsection{Objective Functions}\label{sub:objective}

\subsubsection{WAMS Construction Cost}

The construction cost of WAMS, $C^\text{WAMS}$, can be defined as a combination of the costs for all components involved, i.e., 
\begin{align}\label{eqn:cwams}
C^\text{WAMS}=&\sum_{i\in\mathcal{V}}C^\text{PMU}_i m_i^\prime + \sum_{i\in\mathcal{V}}\sum_{j\in\mathcal{V}}C^\text{DULR}_{ij} d_{ij} \notag\\
+&\sum_{i\in\mathcal{V}}C^\text{PDC}_i p_i + \sum_{k=1}^K C^\text{MS}_k u_k,
\end{align}
where $C^\text{PMU}_i$, $C^\text{DULR}_{ij}$, and $C^\text{PDC}_i$, are the installation costs of the PMU, DULR, and PDC installed at bus $i$, respectively, and $C^\text{MS}_k$ is the disruption cost of substation $k$. $\sum_{j\in\mathcal{V}}d_{ij}$ corresponds to the total number of DULRs installed on branches connected with bus $i$ on the $i$ end and hence $\sum_{i\in\mathcal{V}}\sum_{j\in\mathcal{V}}C^\text{DULR}_{ij} d_{ij}$ determines the cost of installing all DULRs in the system.

\textcolor{blue}{The above objective function presents a general cost model for WAMS construction.
However, in practice, utilities may take advantage of existing substation work that already receives an outage to install devices.
This can be reflected in the model by setting the corresponding substation interruption cost to zero.
In other words, installing PMUs and DULRs in the substation does not incur additional interruption cost and the optimization problem will favor the substation due to achieving lower total cost.}

\subsubsection{Overall System Unreliability Index}

As defined in Section \ref{sub:reliability}, WAMS unreliability index $U^\text{WAMS}$ gives the degree of unreliability subject to all system states. A larger $U^\text{WAMS}$ means the system has a higher possibility of becoming incompletely observable with line outages. Consequently, the system reliability can be improved by minimizing $U^\text{WAMS}$ as an objective function in the proposed multi-objective WPP.

\subsubsection{Total WAMS Data Traffic}

Based on our communication network model, the maximum system data traffic rate received by PDC on bus $j$ can be formulated as $\sum_{i\in\mathcal{V}}(|\mathcal{N}_i+1|)l_{ij}$. We can determine the system data traffic by aggregating the rates of all measurement devices and PDCs. 
%However, as $\mathcal{N}_i^\prime$ is inferred from $\mathbf{M}$ and $\mathbf{D}$, \textcolor{blue}{$\sum_{i\in\mathcal{V}}(|\mathcal{N}_i^\prime+1|)l_{ij}$ is an integer quadratic polynomial and it can drastically increase the computational complexity of the resultant optimization problem.}  \textcolor{red}{(this can be arguable)}. \textcolor{blue}{Therefore in our formulation}, instead of optimizing the actual data traffic in the system, we try to minimize the maximum possible system data traffic. \textcolor{blue}{This can be achieved by} replacing $\mathcal{N}_i^\prime$ with $\mathcal{N}_i$ \textcolor{red}{(can we simply say we minimize the maximum possible system data traffic without mentioning the actual traffic?)}. As $\mathcal{N}_i$ is network topology dependent only, 
Thus, the maximum system data traffic rate $D^\text{WAMS}$ is given by:
\begin{equation} \label{dwams}
D^\text{WAMS}=\sum_{i\in\mathcal{V}}\sum_{j\in\mathcal{V}}(q_{ij}+q_{j c}\eta)LF(|\mathcal{N}_i|+1)l_{ij}
\end{equation}
where $c$ is the location of the central controller, rendering $H_i^\text{PDC}=q_{ic}$.
%\textcolor{blue}{\eqref{dwams} is linear.}
%In this formulation, $\sum_{j\in\mathcal{V}}m_{ij}$ stands for the total number of synchrophasors the PMU on bus $i$ generates, which corresponds to the $|\mathcal{N}_i^\prime|+1$ value introduced in Section \ref{sub:communication}. Similarly, $\sum_{i\in\mathcal{V}}\sum_{j\in\mathcal{V}}d_{ij}$ is the total number of DULRs installed in the system, each of which generates two synchrophasors at one time. Consequently, $\sum_{i\in\mathcal{V}}\sum_{j\in\mathcal{V}}(m_{ij}+2d_{ij})l_{ij^\prime}LF$ is the total data traffic rate received by PDC on $j^\prime$. 
Multiplying the maximum data rate from bus $i$ with $q_{ij}LF$ gives its data traffic to the PDC and the term $q_{jc}\eta LF$ specifies the compressed data traffic from PDC to the central controller. The total data traffic in WAMS is developed by aggregating these two components as given in \eqref{dwams}.

\subsection{Constraints}

%Control variable and practical constraints shall be imposed on the proposed WPP to generate feasible WAMS planning solutions, which are elaborated on in this section.
There are a number of constraints governing the control and ancillary variables.

\subsubsection{Control Variable Constraints}\label{sss:variable_constraint}

We introduce several constraints to define the characteristics of the control variables $\mathbf{M}$, $\mathbf{D}$, $\mathbf{M}^\prime$, $\mathbf{D}^\prime$, and $\mathbf{L}$:
\begin{align}
m_{ij} = 0, &\text{ }\forall(i,j)\notin\mathcal{E} \text{ and } i\neq j \label{con:m}\\
d_{ij} = 0, &\text{ }\forall(i,j)\notin\mathcal{E}\label{con:d}\\
m_i^\prime\geq m_{ij}, &\text{ }\forall i,j\in\mathcal{V} \label{eqn:m_i_prime}\\
d_i^\prime\geq m_i^\prime, d_i^\prime\geq d_{ij}&\text{ }\forall i,j\in\mathcal{V} \label{eqn:d_i_prime}\\
l_{ij} \leq p_j, &\text{ }\forall i,j\in\mathcal{E}\label{con:l1}\\
\sum_{j\in\mathcal{V}}l_{ij}=d_i^\prime, &\text{ }\forall i\in\mathcal{E}\label{con:l2}.
\end{align}
\eqref{con:m} imposes that PMUs can only measure the current phasors of existing branches in the system, and \eqref{con:d} confines that DULRs can only be installed on existing branches. Meanwhile, based on the definitions of $\mathbf{M}$ and $\mathbf{D}$ given in Section \ref{sub:variable}, $m_{ii}=1$ if a PMU is installed on bus $i$, and $d_{ij}=1$ if a DULR is installed on branch $(i,j)$ at the $i$ end. In such cases, we have $m_i^\prime=1$ and $d_i^\prime=1$ according to \eqref{eqn:m_i_prime} and \eqref{eqn:d_i_prime}, respectively. Finally, synchrophasors generated by measurement devices should be sent to PDCs installed at some buses and this gives \eqref{con:l1}. Moreover, as each measurement device should report to one PDC, and buses without measurement devices should be excluded from data traffic calculation, we have \eqref{con:l2}.

\subsubsection{Bus Observability Constraint}

According to \eqref{eqn:ui}, $o_{i,s}$'s are required to calculate $U_i$. As each bus can be observed by either PMUs or DULRs, $o_{i,s}$ is constrained by:
\begin{equation}\label{eqn:observability}
o_{i,s}\leq\sum_{j\in\mathcal{V}}a_{ij,s}(m_{ji} + d_{ji}) + d_{ij}, \forall i\in\mathcal{V}.
\end{equation}
In \eqref{eqn:observability}, $a_{ij,s}m_{ji}$ determines whether a PMU is installed on bus $j$ and measures the current phasor of branch $(j,i)$, rendering bus $i$ observable. $a_{ij,s}m_{ji}$ also includes the cases with $i=j$, where PMU installed on $i$ directly measures the voltage phasor of $i$. Similarly, $a_{ij,s}d_{ji}$ describes the observability of bus $i$ by DULR on branch $(j,i)$ at the $j$ end. Minimizing $U^\text{WAMS}$ is equivalent to maximizing all $o_{i,s}$'s based on \eqref{eqn:observability} as $U^\text{WAMS}=\sum_{i\in\mathcal{V}}\sum_{s\in S}p_s(1-o_{i,s})$, according to \eqref{eqn:ui} and \eqref{eqn:uwams}. Thus, when bus $i$ is observable, the right-hand side of \eqref{eqn:observability} is greater than or equal to one, resulting in $o_{i,s}=1$.

\subsubsection{Redundant Observability Constraint}\label{sss:observability}

WAMS needs to ensure system observability. While the unreliability objective can improve the WAMS reliability subject to transmission line outage failures, it is more practical to provide $N-t$ measurement redundancy to the critical buses to counteract both line outages and measurement device failures, where $t$ is the measurement redundancy degree. At least $t+1$ devices should ensure observability to each critical bus to support $N-t$ measurement redundancy. Other non-critical buses should provided with full observability under $s_0$. We can guarantee $N-t$ observability for critical buses and full observability for non-critical buses with the following:
\begin{equation}
\sum_{j\in\mathcal{V}}[a_{ij}(m_{ji}+d_{ji})+d_{ij}] \geq t+1,\forall i\in\mathcal{V}.
\end{equation}
$\sum_{j\in\mathcal{V}}a_{ij}(m_{ji}+d_{ji})$ determines if bus $i$ can be observed by the PMUs, or DULRs on branch $(j,i)$ at the $j$ end. $\sum_{j\in\mathcal{V}}d_{ij}$ is the total number of DULRs installed on branches connecting to bus $i$ at the $i$ end, which can all provide voltage phasor measurement for $i$.  The summation of these two terms gives the total number of measurement devices observing $i$, which is greater than $t$ to maintain $N-t$ observability for the critical buses. For non-critical buses, $t$ is assigned with zero to guarantee basic full observability. There exists some work about identifying critical buses in a power system, e.g., \cite{pal_pmu_2014}, and we may utilize the results thereafter  to determine $t$ for each bus in the system.

\subsubsection{PMU Measurement Channel Constraint}

PMUs are provided with a limited number of measurement channels for data communications. It is inevitable that some branches cannot be observed even though the adjacent buses have PMUs installed. Hence, to be more practical, we introduce the following constraint:
\begin{equation}
\sum_{j\in\mathcal{V}\setminus i}m_{ij}\leq T^\text{PMU},\text{ }\forall i\in\mathcal{V},
\end{equation}
where $T^\text{PMU}$ is the maximum allowed number of current phasor measurement channel for PMU.

\subsubsection{Substation Interruption Constraint}\label{sss:substation}

When a PMU or DULR is installed, the involved substation needs to be interrupted. Here we employ $\mathbf{U}$ to indicate whether each of the substations  needs to be interrupted and this is illustrated with the following:
\begin{equation}\label{con:u}
u_k \geq m_{ij}\text{ and }u_k \geq d_{ij}, \forall i\in\mathcal{S}_k,j\in\mathcal{V},k=1,\cdots,K.
\end{equation}
As illustrated in Section \ref{sss:variable_constraint}, $m_{ii}$ or $d_{ij}$ is set to one if a PMU or DULR is installed on bus $i$ in substation $k$. In such cases, the corresponding substation needs to be interrupted, making $u_k=1$ by \eqref{con:u}. 

Constraints \eqref{con:m}--\eqref{con:u} describe a set of mandatory constraints on the control variables and WAMS observability requirement. Besides,  there are also some constraints corresponding to practical WAMS requirements.

\subsubsection{Prohibited Substations}\label{sss:prohibited}

In practice, some substations in the power system may not be interrupted at all due to various reasons. For example, substations serving critical loads may undermine the power network security if disrupted. It can also be difficult to install measurement devices on rural substations where no proper CI has been established. In such cases, the installation of PMU and DULR will be prohibited by introducing the following constraint:
\begin{equation}
m_{ij}=0\text{ and }d_{ij}=0, \forall i\in\mathcal{V}^\text{prohibit},j\in\mathcal{V},
\end{equation}
where $\mathcal{V}^\text{prohibit}$ is the set of buses located in these prohibited substations.

\subsubsection{Existing PMU Devices}\label{sss:existing}

In some situations, some substations may have pre-installed PMUs, which can be integrated into WAMS. An additional constraint is imposed to describe the observability introduced by these PMUs:
\begin{equation}
m_{ij}=1,\forall i\in\mathcal{V}^\text{exist}, j\in\mathcal{V}_i^\text{observe},
\end{equation}
where $\mathcal{V}^\text{exist}$ is the set of buses with PMUs pre-installed. For each bus $i$ in $\mathcal{V}^\text{exist}$, $\mathcal{V}_i^\text{observe}$ is the set of other observed buses connected to $i$ by those branches whose current phasors are measurable. Moreover, the cost of these PMUs should be removed from $C^\text{WAMS}$, as originally \eqref{eqn:cwams} includes the cost of all PMUs in the system. This can be achieved by modifying \eqref{eqn:m_i_prime} to
\begin{equation}\label{eqn:m_i_prime_new}
m_i^\prime\geq m_{ij}, \forall i\in\mathcal{V}\setminus\mathcal{V}^\text{exist}, j\in\mathcal{V}\text{ and }m_i^\prime=0, \forall i\in\mathcal{V}^\text{exist}.
\end{equation}
as $m_i^\prime$ is involved in \eqref{eqn:cwams} to calculate the construction cost. \eqref{eqn:m_i_prime_new} excludes the existing PMUs by setting their respective $m_i^\prime$ values to zero, thus making \eqref{eqn:cwams} accurate.

\subsection{Multi-objective Optimization Problem}\label{sub:problem}

Utilizing previously defined objective functions and constraints, we formulate a multi-objective optimization problem as follows:
\begin{align}\label{eqn:multiobjective}
\underset{\mathbf{M},\mathbf{D},\mathbf{P},\mathbf{L},\mathbf{M}^\prime,\mathbf{D}^\prime,\mathbf{U}}{\textrm{minimize}} & (C^\text{WAMS},U^\text{WAMS},D^\text{WAMS})\notag\\
\textrm{subject to}\hspace{1.5em} & \eqref{con:m}-\eqref{eqn:m_i_prime_new}\notag
\end{align}
This optimization problem will develop the optimal WAMS plan based on these conflicting objectives; minimizing WAMS unreliability results in installing measurement devices at all possible positions, which in turn drastically increases the total construction cost; minimizing the total data traffic requires equipping all measurement devices with PDCs for data compression, but this increases the cost. Therefore, a Pareto frontier will be developed for decision-making.

\subsection{Discussions}

\textcolor{blue}{The multi-objective optimization problem defined above %in Section \ref{sub:problem} 
provides solutions to a generalized WAMS construction problem considering different constructional objectives.
The construction deduced from the solutions may take place over a specified period of time.
As such, there may be a transitional period in which WAMS may work together with the conventional SCADA system before being fully deployed.
The operational issues have been addressed in previous work (see \cite{gol_hybrid_2015,kashyap_power_2014} for examples).
In addition, some previous research suggests to place PMU over SCADA, e.g., \cite{li_framework_2014,glazunova_pmu_2009}.
Typically this can be achieved by including additional constraints for SCADA in the conventional PMU placement problem.
Our formulated multi-objective optimization problem can also adopt these constraints with ease, and the resulting problem can handle the WAMS construction problem considering an existing SCADA system.}

\textcolor{blue}{Sometimes utilities may impose a bandwidth limit for WAMS data transmission instead of minimizing the data traffic.
This can be addressed by transforming the corresponding objective function(s) into constraints.
For example, suppose that the total data traffic from bus $i$ to PDC at bus $j$ is $\overline{D}_{ij}$.
The data traffic objective \eqref{eqn:cwams} can be transformed into a series of constraints, each of which limits the bus-to-PDC data traffic as follows:
\begin{equation}
  LF(|\mathcal{N}_i|+1)l_{ij}\leq\overline{D}_{ij},\forall i,j\in\mathcal{V}.
\end{equation}
The constraints for maximum construction cost (budget) and minimum reliability index can be transformed from \eqref{eqn:cwams} and \eqref{eqn:uwams} in a similar way, respectively.}

\textcolor{blue}{Last but not least, besides being applicable to new power systems without existing WAMS, the proposed formulation can also generate optimal construction plans in the presence of existing measurement devices, which is common in some developed grids, e.g., North America systems \cite{_pmus_????}.
In such systems, the utilities may need to gradually introduce new devices to the existing WAMS.
The proposed formulation can still develop optimal construction plans to achieve objectives such as minimizing construction cost and maximizing post-construction reliability.}

\section{Case Studies and Discussions}

\begin{figure}
\includegraphics[width=\linewidth]{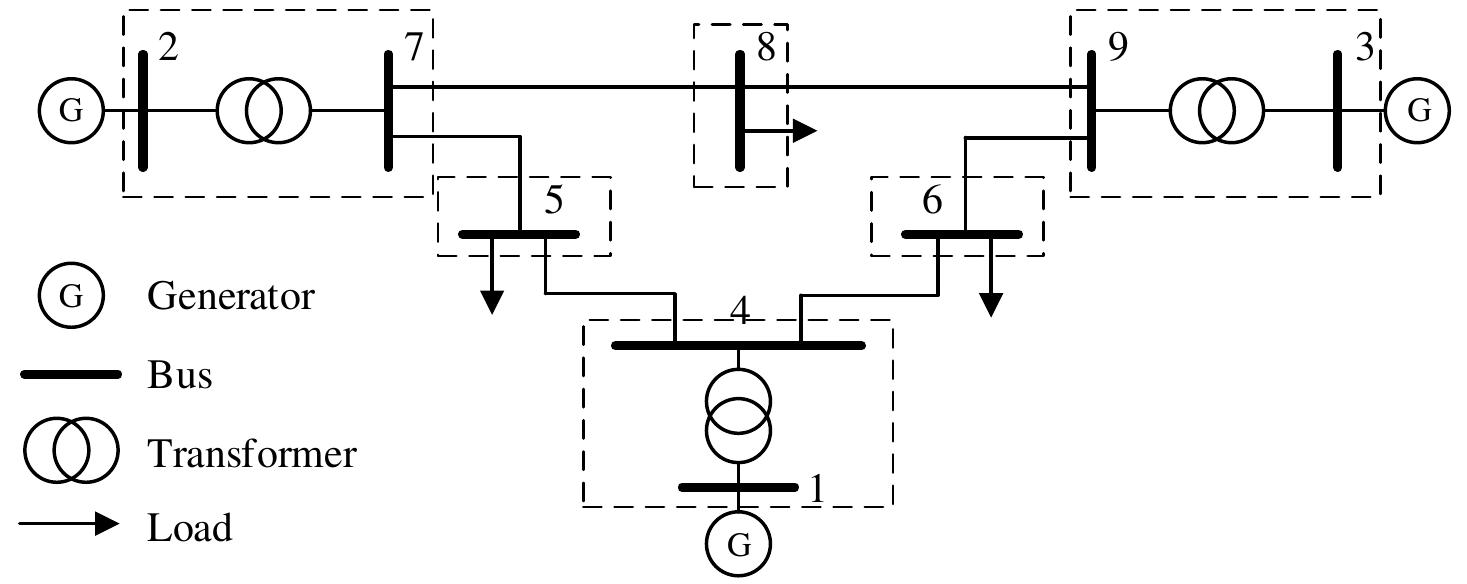}
\caption{Line diagram of the IEEE 9-bus system and its substation setting. Dashed boxes are substations with one or multiple buses.}
\label{fig:9bus}
\end{figure}

\begin{figure*}
\centering
\subfloat[IEEE 9-bus system]{
\centering
\includegraphics[width=0.45\linewidth]{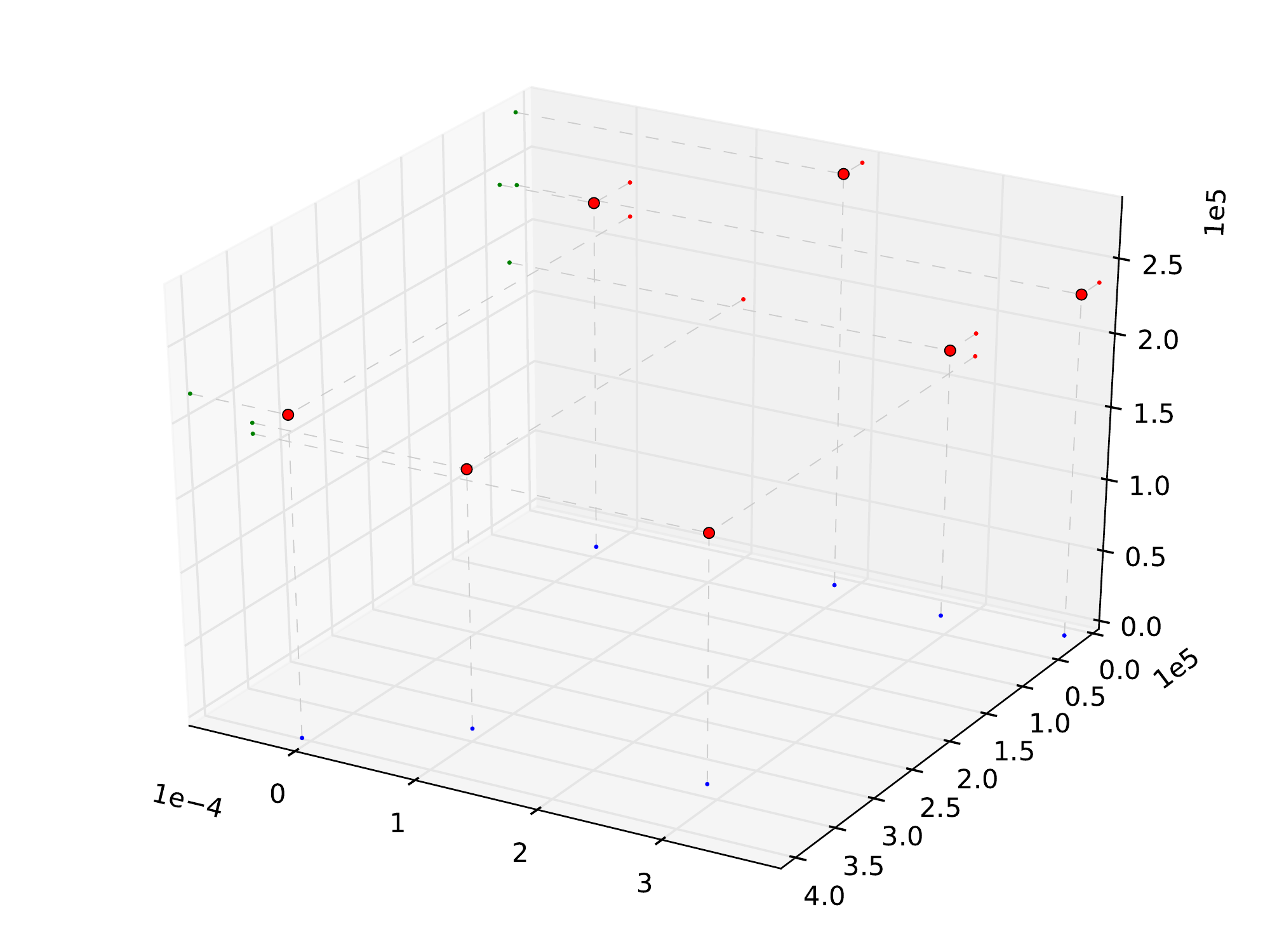}
}
\subfloat[IEEE 57-bus system]{
\centering
\includegraphics[width=0.45\linewidth]{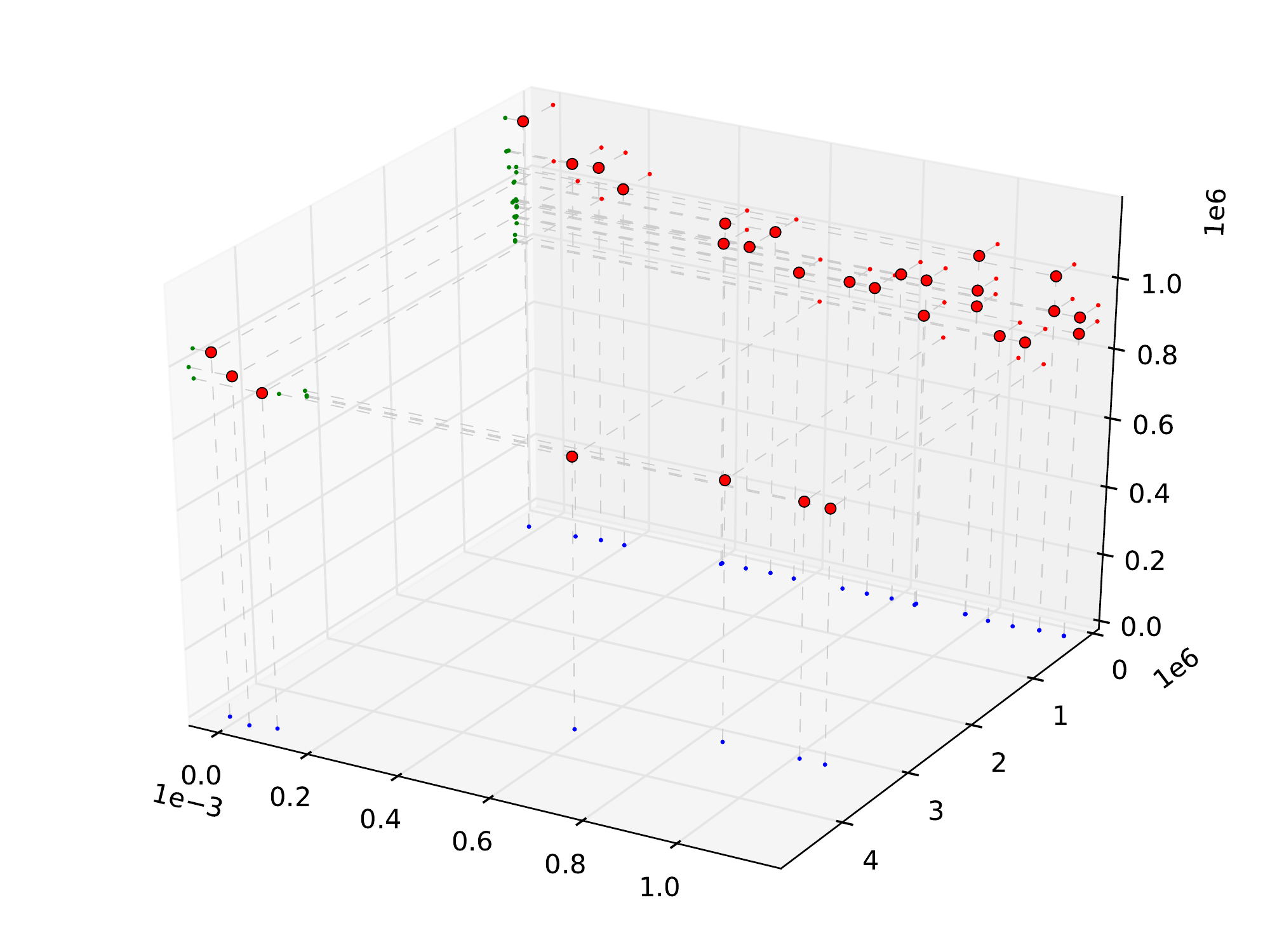}
}
\caption{Pareto optimal performance of (a) IEEE 9-bus system and (b) IEEE 57-bus system.}
\label{fig:pareto}
\end{figure*}

To assess the efficacy of the proposed framework, we employ the IEEE 9-bus and 57-bus systems for performance evaluation. The system parameters are set according to the studies reported by various utilities. Specifically, all costs are developed from quotations of General Electric Grid Solutions \cite{_ge_????} and the U.S. Department of Energy PMU installation cost report \cite{_factors_2014}, and the data frame information is derived from \cite{_ieee_2011}. All parameter values are listed in Table \ref{tbl:param}.

The test systems are modified to simulate substation interruptions. Buses connected with transformers are considered to be located in the same substation. Other buses without any transformers attached form independent substations. See Fig. \ref{fig:9bus} as an illustration of the modified IEEE 9-bus system, where the dashed boxes are substations. As a result, the IEEE 9-bus and 57-bus  systems have six and 42 substations, respectively. In addition, we assume that the control centers are located at bus 8 and 38 for the two test systems, respectively, which are arbitrarily selected and their locations do not have significant impacts on the system performance. 
\textcolor{blue}{For simplicity, we assume that CI shares the same topology with the power network.
This assumption only influences the value of the distance matrix $\mathbf{Q}$, which is a constant.
Other topologies, e.g., star and ring networks, can be easily adopted by changing the value of $\mathbf{Q}$.}

\begin{table}
\scriptsize
\centering
\caption{Parameter Settings}
\label{tbl:param}
\begin{tabular}{cccc}
  \hline
  $C^\text{PMU}_i$ \cite{_ge_????} & $C^\text{DULR}_{ij}$ \cite{_ge_????} & $C^\text{PDC}_{i}$ \cite{_ge_????} & $C^\text{MS}_{k}$ \cite{pal_pmu_2017,_factors_2014}\\ \hline
  \$8819.46 & \$5146.87 & \$7750.00 & \$40000.0 \\
  \hline
  \hline
  $p_l$ \cite{gomez_reliability_2015,wang_reliability-based_2014} & $\eta$ \cite{zhang_application_2015} & $LF$ \cite{_ieee_2011} & $T^\text{PMU}$ \cite{_ge_????} \\ \hline
  0.99 & 0.0877 & 20160 bps & 2 \\
  \hline
\end{tabular}
\end{table}

\begin{table*}
\centering
\scriptsize
\caption{Pareto Optimal Solutions for IEEE 9-bus System}
\label{tbl:IEEE_9}
\begin{tabular}{c|rrr|ccc}
  \hline
  \multirow{2}*{Solution} & \multicolumn{3}{c|}{Objective Values} & \multirow{2}*{PMUs} & \multirow{2}*{DULRs} & \multirow{2}*{PDCs} \\\cline{2-4}
  & $C^\text{WAMS}$ & $U^\text{WAMS}$ & $D^\text{WAMS}$ & & & \\
  \hline
1 & 1.70E+05 & 2.88E-04 & 3.44E+05 & 1(1)$\rightarrow$9, 2(2)$\rightarrow$9, 3(3)$\rightarrow$9 & 4(6)$\rightarrow$9, 7(5)$\rightarrow$9, 9(8)$\rightarrow$9 & 9 \\
2 & 1.77E+05 & 9.61E-05 & 3.44E+05 & 1(1)$\rightarrow$9, 2(2)$\rightarrow$9, 3(3)$\rightarrow$9, 7(5,7,8)$\rightarrow$9, 9(6,8,9)$\rightarrow$9 & 4(6)$\rightarrow$9 & 9 \\
3 & 1.85E+05 & 2.88E-04 & 3.54E+04 & 1(1)$\rightarrow$1, 2(2)$\rightarrow$2, 3(3)$\rightarrow$3 & 4(6)$\rightarrow$1, 7(5)$\rightarrow$2, 9(8)$\rightarrow$3 & 1,2,3 \\
4 & 2.18E+05 & 0.00E+00 & 4.10E+05 & 1(1)$\rightarrow$2, 2(2)$\rightarrow$2, 3(3)$\rightarrow$2, 9(6,8,9)$\rightarrow$2 & 4(6)$\rightarrow$2, 5(7)$\rightarrow$2, 7(8)$\rightarrow$2 & 2 \\
5 & 2.35E+05 & 3.84E-04 & 2.48E+04 & 1(1)$\rightarrow$1, 2(2)$\rightarrow$2, 3(3)$\rightarrow$3, 6(4,6,9)$\rightarrow$6, 7(5,7,8)$\rightarrow$2 & N/A & 1,2,3,6 \\
6 & 2.42E+05 & 0.00E+00 & 4.60E+04 & 1(1)$\rightarrow$4, 2(2)$\rightarrow$7, 3(3)$\rightarrow$9, 9(6,8,9)$\rightarrow$9 & 4(6)$\rightarrow$4, 5(4)$\rightarrow$5, 7(8)$\rightarrow$7 & 4,5,7,9 \\
7 & 2.84E+05 & 1.92E-04 & 2.48E+04 & 1(1)$\rightarrow$4, 2(2)$\rightarrow$7, 3(3)$\rightarrow$9, 6(4,6,9)$\rightarrow$6 & 7(5)$\rightarrow$7, 8(9)$\rightarrow$8 & 4,6,7,8,9 \\
  \hline
\end{tabular}
\end{table*}

The problem is a multi-objective integer linear program (MOILP) and it can be effectively solved by addressing its equivalent bounded weighted-sum program \cite{alves_review_2007}. Each singleton ILP is solved using Gurobi \cite{_gurobi_????}, a high-efficiency numerical optimization solver. All tests are performed on a PC with an Intel Core-i7 CPU at 3.6GHz and 32GB RAM. Problem formulation and simulation scripts are coded with Python.

\subsection{Pareto Optimal Solutions}\label{sub:simu1}

We perform a series of simulation on both IEEE 9-bus and 57-bus systems. The Pareto frontiers are presented in Fig. \ref{fig:pareto}, where IEEE 9-bus and 57-bus systems have seven and 30 Pareto optimal solutions, respectively. In both sub-plots, the x-axis corresponds to $R^\text{WAMS}$, y-axis to $D^\text{WAMS}$, and z-axis to $C^\text{WAMS}$. The objective function values of all Pareto optimal solutions are plotted as the red dots in the 3-D space. To distinguish their relative positions in the space, we also give their projections on x-y, y-z, and x-z 2-D planes, with dashed lines as auxiliary lines.

Although the Pareto frontier is not smooth due to the combinatorial nature of the problem, it contains all the non-dominating solutions. The combinatorial nature of the proposed optimization problem makes these solutions not well-positioned. Changing one value in the control variables can potentially lead to drastic changes in the objective values, resulting in a non-smooth solution space with numerous local optimums.

To better investigate the characteristics of the Pareto optimal solution, the detailed WAMS construction plans for the IEEE 9-bus system are given in Table \ref{tbl:IEEE_9} with their respective objective function values. In the ``PMUs'' column, all PMUs installed are presented in the form of $i(a,b,\allowbreak\cdots)\rightarrow j$, which means that a PMU is installed on bus $i$ and transmits synchrophasors to the PDC located at bus $j$. This PMU makes buses $a,b,\cdots$ observable by measuring the current phasors of their respective connecting branches. Note that the PMU also measures the voltage phasor of its associated bus. ``DULRs'' column also has a similar $i(a)\rightarrow j$ pattern, where a DULR is installed on branch $(i,a)$ on the $i$ end, and sends data to $j$.
The table shows insight on the relationships between objective function values and their corresponding WAMS plans. Solutions 1 and 2 share an identical PMU and DULR allocation scheme, and their $U^\text{WAMS}$ are the same. This is also the case for solutions 6 and 7. Meanwhile, the total number of PDCs installed in the system has a significant impact on $D^\text{WAMS}$ performance, while the measurement device allocation is relatively less influential.

\begin{figure*}
\centering
\subfloat[Optimize $C^\text{WAMS}$ and $U^\text{WAMS}$]{
\centering
\includegraphics[width=0.3\linewidth]{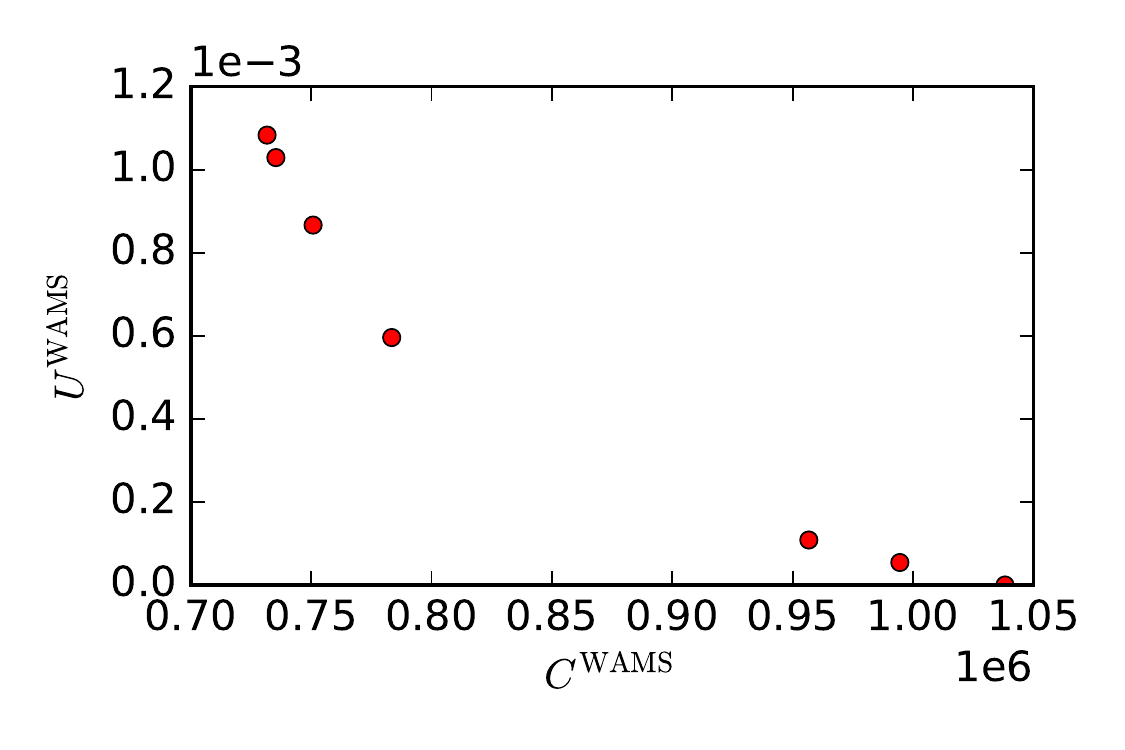}
}
\subfloat[Optimize $C^\text{WAMS}$ and $D^\text{WAMS}$]{
\centering
\includegraphics[width=0.3\linewidth]{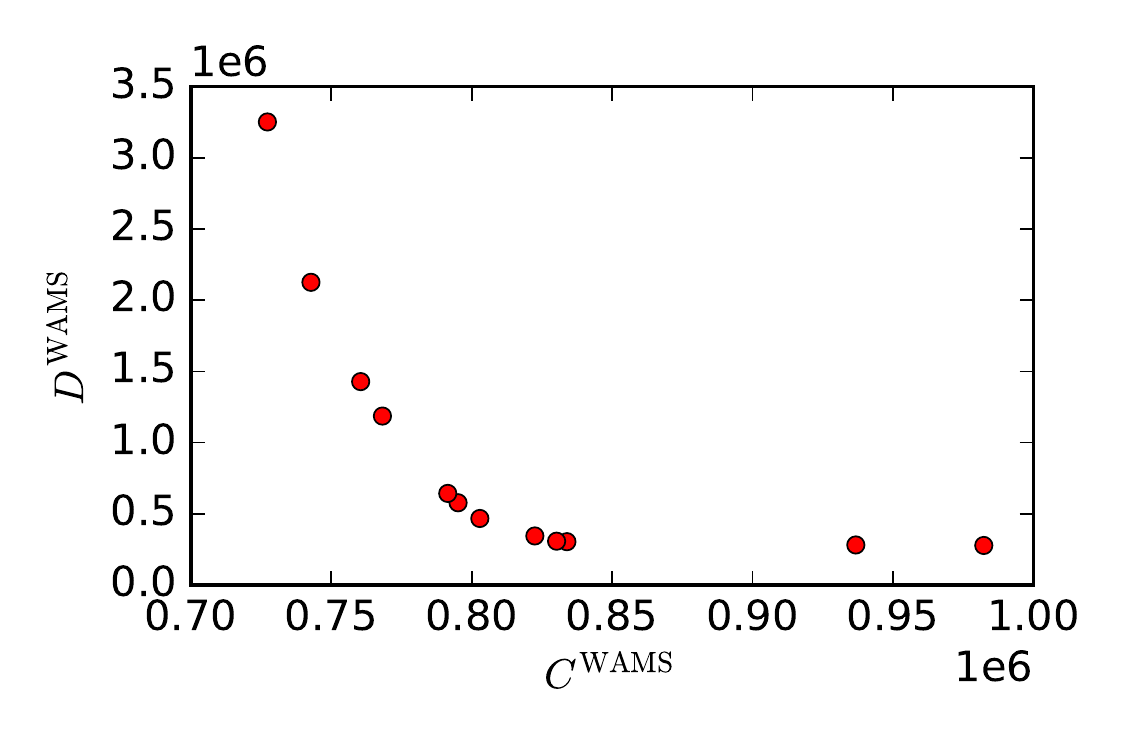}
}
\subfloat[Optimize $U^\text{WAMS}$ and $D^\text{WAMS}$]{
\centering
\includegraphics[width=0.3\linewidth]{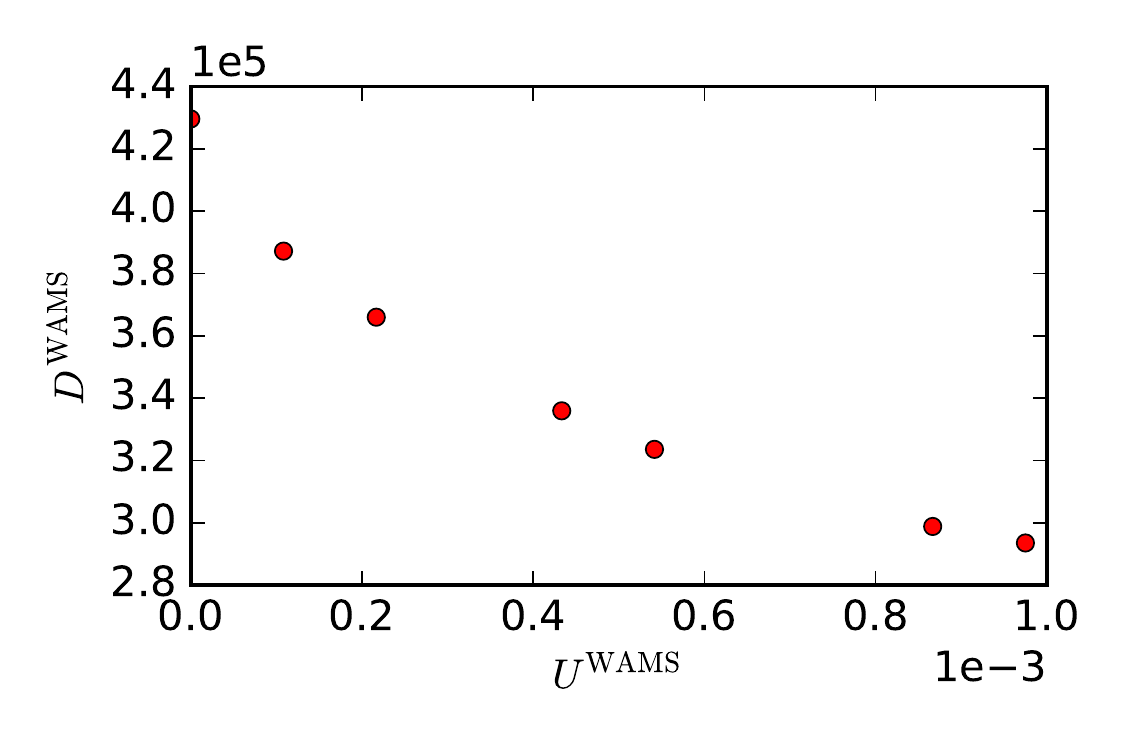}
}
\caption{Pareto optimal performance when jointly optimizing (a) $C^\text{WAMS}$ and $U^\text{WAMS}$, (b) $C^\text{WAMS}$ and $D^\text{WAMS}$, and (c) $U^\text{WAMS}$ and $D^\text{WAMS}$.}
\label{fig:two_objective}
\end{figure*}

We also perform another set of simulation, where we compare two objectives each time to demonstrate their trade-off relationships. We illustrate this test on the IEEE 57-bus system. The Pareto optimal solutions are presented in Fig. \ref{fig:two_objective}. From the plots it is obvious that all three objectives contradict with each other and this contradiction exists in the hypothesis given in Section \ref{sub:problem}. Although the objective functions are different, we can still find some relationships between Figures \ref{fig:pareto}b and \ref{fig:two_objective}. All data points presented in Fig. \ref{fig:two_objective} have a corresponding point in Fig. \ref{fig:pareto}b, while points in the latter figure may become multiple points in the subplots of Fig. \ref{fig:two_objective}. As points in Fig. \ref{fig:two_objective} are non-dominating solutions with respect to two objective functions, they will retain their non-dominating property with the extra objective. Meanwhile, non-dominating solutions in \ref{fig:pareto}b can possibly be non-dominating on all three pairs of objective functions, resulting in multiple points in the 2-D plots.

\textcolor{blue}{Last but not least, we also test the scalability of the proposed optimization problem.
To do this, several large-scale test power systems, namely, IEEE 118-bus, 145-bus, and 300-bus systems, are studied.
We are specifically interested in the tractability of such power systems and the computational time required to produce optimal solutions. The simulation results are presented in Table \ref{tbl:time}.
We list the computational time required to calculate the first optimal solution and all solutions.
It can be observed that the computational time drastically increases with system size.
For all test systems except the 300-bus case, the optimization can be finished within one day.
However, we are unable to develop all Pareto optimal solutions for IEEE 300-bus within reasonable time on the PC set-up.
For very large systems, we may first divide the systems into smaller sub-systems, and then generate WAMS construction plans for them.
Since a smaller problem can be effectively solved, the solutions to all sub-systems can then be combined to construct the sub-optimal solutions for the whole system.
In addition, as power systems are typically sparse, the optimality gap of this approach should be insignificant.
How to integrate the proposed multi-objective optimization with this approach will be investigated in future research.
}

\begin{table}
\scriptsize
\centering
\textcolor{blue}{
\caption{Computation Time Comparison on Different Test Systems}
\label{tbl:time}
\begin{tabular}{ccc}
  \hline
  System & First optimal solution & Total time \\
  \hline
  IEEE 9-bus & 0.04 s & 12.83 s\\
  IEEE 57-bus & 11.25 s & 1348.01 s\\
  IEEE 118-bus & 192.85 s & 22013.62 s\\
  IEEE 145-bus & 473.60 s & 59248.27 s\\
  IEEE 300-bus & 4699.28 s & -\\
  \hline
\end{tabular}
}
\end{table}

\begin{table*}
\centering
\scriptsize
\textcolor{blue}{
\caption{Pareto Optimal Solutions for IEEE 9-bus System with Bus 6 Prohibited from Construction}
\label{tbl:IEEE_9_p6}
\begin{tabular}{c|rrr|ccc}
  \hline
  \multirow{2}*{Solution} & \multicolumn{3}{c|}{Objective Values} & \multirow{2}*{PMUs} & \multirow{2}*{DULRs} & \multirow{2}*{PDCs} \\\cline{2-4}
  & $C^\text{WAMS}$ & $U^\text{WAMS}$ & $D^\text{WAMS}$ & & & \\
  \hline
1 & 1.70E+05 & 2.88E-04 & 3.44E+05 & 1(1)$\rightarrow$7, 2(2)$\rightarrow$7, 3(3)$\rightarrow$7 & 4(6)$\rightarrow$7, 7(5)$\rightarrow$7, 9(8)$\rightarrow$7 & 7 \\
2 & 1.77E+05 & 9.61E-05 & 3.44E+05 & 1(1)$\rightarrow$7, 2(2)$\rightarrow$7, 3(3)$\rightarrow$7, 7(5,7,8)$\rightarrow$7, 9(6,8,9)$\rightarrow$7 & 4(6)$\rightarrow$7 & 7 \\
3 & 1.85E+05 & 2.88E-04 & 3.54E+04 & 1(1)$\rightarrow$1, 2(2)$\rightarrow$2, 3(3)$\rightarrow$3 & 4(6)$\rightarrow$1, 7(5)$\rightarrow$2, 9(8)$\rightarrow$3 & 1,2,3 \\
4 & 1.92E+05 & 9.61E-05 & 3.54E+04 & 1(1)$\rightarrow$4, 2(2)$\rightarrow$7, 3(3)$\rightarrow$9, 7(5,7,8)$\rightarrow$7, 9(6,8,9)$\rightarrow$9 & 4(6)$\rightarrow$4 & 4,7,9 \\
5 & 2.18E+05 & 0.00E+00 & 4.10E+05 & 1(1)$\rightarrow$2, 2(2)$\rightarrow$2, 3(3)$\rightarrow$2, 9(6,8,9)$\rightarrow$2 & 4(6)$\rightarrow$2, 5(4)$\rightarrow$2, 7(8)$\rightarrow$2 & 2 \\
6 & 2.33E+05 & 2.88E-04 & 3.01E+04 & 1(1)$\rightarrow$1, 2(2)$\rightarrow$2, 3(3)$\rightarrow$3 & 4(6)$\rightarrow$1, 7(5)$\rightarrow$2, 8(9)$\rightarrow$8 & 1,2,3,8 \\
7 & 2.35E+05 & 1.03E-02 & 2.48E+04 & 1(1)$\rightarrow$4, 2(2)$\rightarrow$7, 3(3)$\rightarrow$9, 5(4,5,7)$\rightarrow$5, 9(6,8,9)$\rightarrow$9 & N/A & 4,5,7,9 \\
8 & 2.42E+05 & 0.00E+00 & 4.60E+04 & 1(1)$\rightarrow$4, 2(2)$\rightarrow$7, 3(3)$\rightarrow$9, 9(6,8,9)$\rightarrow$9 & 4(6)$\rightarrow$4, 5(4)$\rightarrow$5, 7(8)$\rightarrow$7 & 4,5,7,9 \\
9 & 2.84E+05 & 1.01E-02 & 2.48E+04 & 1(1)$\rightarrow$4, 2(2)$\rightarrow$7, 3(3)$\rightarrow$9, 5(4,5,7)$\rightarrow$5 & 8(7)$\rightarrow$8, 9(6)$\rightarrow$9 & 4,5,7,8,9 \\
  \hline
\end{tabular}
}
\end{table*}

\subsection{Impact of Optional Operational Constraints}

\textcolor{blue}{In the previous test, we investigated the Pareto performance of generalized WAMS construction problem.
As introduced in Sections \ref{sss:prohibited} and \ref{sss:existing}, it is also possible that some substations are reserved from installing measurement devices or already have them installed.
Here we investigate the impact of these practical constraints on the Pareto optimal WAMS construction plans.}

\textcolor{blue}{For simplicity, we focus on IEEE 9-bus system to demonstrate construction results.
We investigate two test cases.
In the first case, we assume that Bus 6 is prohibited from being installed with measurement devices.
In the second one, we additionally assume that Bus 7 is equipped with a PMU.
All other configurations remain the same as stated in Section \ref{sub:simu1}.
The Pareto optimal solutions for these test cases are listed in Tables \ref{tbl:IEEE_9_p6} and \ref{tbl:IEEE_9_p6i7}.}
\begin{table*}
\centering
\scriptsize
\textcolor{blue}{
\caption{Pareto Optimal Solutions for IEEE 9-bus System with Bus 6 Prohibited from Construction and Bus 7 Pre-installed with a PMU}
\label{tbl:IEEE_9_p6i7}
\begin{tabular}{c|rrr|ccc}
  \hline
  \multirow{2}*{Solution} & \multicolumn{3}{c|}{Objective Values} & \multirow{2}*{PMUs} & \multirow{2}*{DULRs} & \multirow{2}*{PDCs} \\\cline{2-4}
  & $C^\text{WAMS}$ & $U^\text{WAMS}$ & $D^\text{WAMS}$ & & & \\
  \hline
1 & 1.25E+05 & 1.92E-04 & 3.44E+05 & 1(1)$\rightarrow$2, 2(2)$\rightarrow$2, 3(3)$\rightarrow$2, 7(5,7,8)$\rightarrow$2 & 4(6)$\rightarrow$2, 9(8)$\rightarrow$2 & 2 \\
2 & 1.28E+05 & 9.61E-05 & 3.44E+05 & 1(1)$\rightarrow$2, 2(2)$\rightarrow$2, 3(3)$\rightarrow$2, 7(5,7,8)$\rightarrow$2, 9(6,8,9)$\rightarrow$2 & 4(6)$\rightarrow$2 & 2 \\
3 & 1.40E+05 & 1.92E-04 & 3.54E+04 & 1(1)$\rightarrow$4, 2(2)$\rightarrow$7, 3(3)$\rightarrow$9, 7(5,7,8)$\rightarrow$7 & 4(6)$\rightarrow$4, 9(6)$\rightarrow$9 & 4,7,9 \\
4 & 1.44E+05 & 9.61E-05 & 3.54E+04 & 1(1)$\rightarrow$4, 2(2)$\rightarrow$7, 3(3)$\rightarrow$9, 7(5,7,8)$\rightarrow$7, 9(6,8,9)$\rightarrow$9 & 4(6)$\rightarrow$4 & 4,7,9 \\
5 & 1.73E+05 & 0.00E+00 & 4.10E+05 & 1(1)$\rightarrow$2, 2(2)$\rightarrow$2, 3(3)$\rightarrow$2, 7(5,7,8)$\rightarrow$2, 9(6,8,9)$\rightarrow$2 & 4(6)$\rightarrow$2, 5(4)$\rightarrow$2 & 2 \\
6 & 1.88E+05 & 2.88E-04 & 3.01E+04 & 1(1)$\rightarrow$1, 2(2)$\rightarrow$2, 3(3)$\rightarrow$3, 7(5,7,8)$\rightarrow$2 & 4(6)$\rightarrow$1, 8(9)$\rightarrow$8 & 1,2,3,8 \\
7 & 1.97E+05 & 0.00E+00 & 4.60E+04 & 1(1)$\rightarrow$4, 2(2)$\rightarrow$7, 3(3)$\rightarrow$9, 7(5,7,8)$\rightarrow$7, 9(6,8,9)$\rightarrow$9 & 4(6)$\rightarrow$4, 5(7)$\rightarrow$5 & 4,5,7,9 \\
  \hline
\end{tabular}
}
\end{table*}

\textcolor{blue}{Comparing with Table \ref{tbl:IEEE_9}, the Pareto optimal solutions in the original test case can be grouped into two classes: solutions with or without constructions at Bus 6.
The former solutions, i.e., Solutions 5 and 7 in Table \ref{tbl:IEEE_9} are replaced by Solutions 4, 6, 7, and 9 in Table \ref{tbl:IEEE_9_p6}.
This is because introducing a new constraint changes the solution space of the problem, leading to different optimal solutions.
On the other hand, the latter solutions maintain their construction plans but the reliability is reduced.
This accords with the nature of system reliability.
The new constraint rules out the possibility of observing Bus 6 by installing measurement devices on the bus.
Therefore, fewer devices can observe the bus and its robustness against contingencies is undermined.}

\textcolor{blue}{When there is a pre-installed PMU at Bus 7, all solutions in Table \ref{tbl:IEEE_9_p6} which plan to install devices on the bus is preserved.
The remaining two, i.e., Solutions 7 and 9, are removed since they are dominated due to the lower cost induced by other solutions.
To conclude, prohibiting device installation on buses generally decreases the system reliability and having pre-installed devices significantly reduce the system construction cost.}

\subsection{Comparisons with State-of-the-Art}

To further evaluate the proposed framework, we compare our proposed model with some state-of-the-art existing work based on the IEEE 57-bus system. As no other work consider $U^\text{WAMS}$ and $D^\text{WAMS}$, we focus on $C^\text{WAMS}$ for assessment. We compare the optimal solutions developed by the other work with the Pareto optimal solution with the least WAMS construction cost, presented in Table \ref{tbl:cost}. In this table, the cost performance of our proposed WPP is presented with the detailed WAMS construction plan, which is in the form as in Table \ref{tbl:IEEE_9} without data transmission destinations. In addition, we also study the maximum number of PMU channels needed, as it is a major concern in real-world PMU installation process. Last but not least, we consider the impact introduced by unknown transformer tap ratios to the system observability, which is listed under the ``Unknown Tap'' column.
In this comparison, Immunity Genetic Algorithm (IGA) \cite{aminifar_optimal_2009}, Integer Programming (IP) approach \cite{aminifar_contingency-constrained_2010}, and Cellular Learning Automata \cite{mazhari_multi-objective_2013} are employed. We employ the same substation interruption cost and PMU installation cost as in Table \ref{tbl:param}, and set the cost for PMUs with more CTs to $\$12530.00$ \cite{_ge_????}. For observability, the PMUs installed in these plans are assumed to observe all connecting branches.

From Table \ref{tbl:cost} we can see that the construction cost of the proposed framework does not perform as good as the others. However, none of the compared algorithms can actually achieve full numerical observability on the test system due to unknown transformer tap ratios. This incomplete observability may contribute to their low WAMS construction cost. For these plans to achieve full observability in absence of tap ratios, extra measurement devices are required, which will incur extra cost \cite{aminifar_contingency-constrained_2010}. Another possible reason of their low cost is that these solutions introduces zero-injection buses to improve the observability. As an integer linear programming formulation of zero-injection buses was proposed in \cite{pal_pmu_2017}, it can be employed in our framework with minimal effort to further improve the performance.
Considering these analyses, our proposed WPP can generate a feasible WAMS plan providing full numerical observability with reasonable construction cost.

\begin{table*}
\begin{threeparttable}
\centering
\scriptsize
\caption{Cost Comparison of the Proposed Framework with State of the Art For IEEE 57-bus System}
\label{tbl:cost}
\begin{tabular}{l|r|l|cc}
  \hline
  \multicolumn{1}{c|}{\multirow{2}*{Method}} & \multicolumn{1}{c|}{\multirow{2}*{Cost}} & \multicolumn{1}{c|}{\multirow{2}*{Measurement Devices}} & \multicolumn{2}{c}{Constraints} \\\cline{4-5}
  & & & $\max{\{T^\text{PMU}\}}$ & Unknown Tap \\
  \hline
  \multirow{4}*{Proposed} & \multirow{4}*{7.24E+05\tnote{a}} & PMUs: 4(5,6), 11(9,13), 12(16,17), 15(1,14), 29(28,52), 32(31,33), 37(38,39), & \multirow{4}*{\textbf{2}} & \multirow{4}*{\textbf{Observable}} \\
   & & 47(46,48), 50(49,51), 54(53,55), 56(42,57) & & \\
   & & DULRs:3(2), 7(8), 12(10), 18(19), 20(19), 21(22),
 24(23), 25(30), 26(27), 34(35), & & \\
   & & 40(36), 43(41), 45(44) & & \\\hline
  IGA \cite{aminifar_optimal_2009}\tnote{b} & \textbf{5.78E+05}\tnote{c} & PMUs: 1, 6, 13, 19, 25, 29, 32, 38, 51, 54, 56 & 6 & Unobservable \\\hline
  \multirow{2}*{IP \cite{aminifar_contingency-constrained_2010}\tnote{b,d}} & \textbf{5.78E+05}\tnote{c} & PMUs: 1, 4, 13, 20, 25, 29, 32, 38, 51, 54, 56 & 6 & Unobservable \\\cline{2-5}
   & 9.98E+05 & PMUs: 1, 2, 6, 12, 14, 19, 21, 27, 29, 30, 32, 33, 41, 44, 49, 51, 53, 55, 56 & 5 & Unobservable \\\hline
  CLA \cite{mazhari_multi-objective_2013}\tnote{b} & \textbf{5.78E+05}\tnote{c} & PMUs: 1, 4, 13, 20, 25, 29, 32, 38, 51, 54, 56 & 6 & Unobservable \\
  \hline
\end{tabular}
\begin{tablenotes}
\item[a] For a fair comparison, the PDC cost is removed to keep accordance with other methods.
\item[b] Installed PMUs are assumed to be able to observe all connecting branches.
\item[c] This minimal cost cannot guarantee system numerical observability without known transformer tap ratio.
\item[d] Multiple optimal solutions are given subject to different constraints. Two most cost-efficient ones are listed for comparison.
\end{tablenotes}
\end{threeparttable}
\end{table*}

\section{Conclusion}

In this paper we propose a unified framework for constructing the future WAMS based on several realistic considerations. To be specific, we formulate a practical cost model for WAMS construction, considering the fact that the measurement devices should be installed at substations, whose interruption cost should not be ignored during installation. In addition, no real-time transformer tap ratio information is required and this fits into the practical situations better in the power system. Moreover, we allocate PMUs to different buses, and at the same time assign branches to the PMUs for measurement.

In the proposed model, we consider three representative WAMS construction objectives, namely, the construction cost, system reliability, and synchrophasor data traffic. They unify the most common objectives considered in most related work. While different WAMS may stress on distinctive objectives, they should be jointly optimized to develop practical optimal WAMS construction plans. Therefore, we propose a multi-objective WPP for developing multiple Pareto optimal solutions to suit different purposes of the utilities.
%Utilities may select appropriate Pareto optimal solutions as their WAMS construction plans considering their respective WAMS requirements.
We verify the proposal model on the IEEE 9-bus and 57-bus systems. The simulation results indicate that multiple Pareto optimal solutions can be developed. In addition, when maintaining the system numerical observability, our proposed framework can result in optimal WAMS construction plans with minimal cost.
This work develops a comprehensive framework for most practical WAMS construction designs.

\section*{Acknowledgement}

This work was supported by the Theme-based Research Scheme of the Research Grants Council of Hong Kong, under Grant No. T23-701/14-N.

\bibliographystyle{elsarticle-num}
\bibliography{zotero}

\end{document}